\documentclass[12pt,a4paper]{article}
\usepackage[english]{babel}
\usepackage{amsmath,amsfonts,amssymb,amsthm}
\theoremstyle{plain} \numberwithin{equation}{section}

\newtheorem{lemma}{Lemma}[section]

\theoremstyle{definition}
\theoremstyle{remark}

\begin{document}

\title{ Dynamics of Bloch Electrons in Time Dependent External Electric
Fields: Bounds for Interband Transitions}
\author{  A. Nenciu\\
Faculty of Applied Sciences University ``Politehnica'' of
Bucharest,\\ Splaiul Independentei 313, RO-060042 Bucharest,
Romania}

\date{}
\maketitle

\begin{abstract}
Using adiabatic expansions formalism, upper bounds for interband
transitions for Bloch electrons in slowly varying in time electric
fields are obtained. These bounds imply the validity of one-band
approximation on long time scales.

\end{abstract}

\section{Introduction}
This paper is devoted to the generalization of the main result in
\cite{NN1} concerning the smallness of the interband transitions
for homogeneous time independent external electric fields to
slowly time dependent electric fields. The study of Bloch
electrons in a time independent electric field has a long and
distinguished history. The subject is as old as the quantum theory
of solids (see e.g. \cite{N} for an extensive discussion) but, as
the problem of the interband transitions is concerned, the real
story started with the papers of Wannier \cite{W1}, \cite{W2} who
argued that in the presence of a weak homogeneous time independent
electric field the energy bands of the crystal are "deformed" and
there are no interband transitions between the deformed bands.
Moreover, the Hamiltonian restricted to a simple deformed band
consists of a ladder of discrete eigenvalues (Stark-Wannier
ladder). Wannier claims were challenged by Zak \cite{Z1} on the
ground that in the presence of arbitrarily weak field the spectrum
becomes continuous so Stark-Wannier ladders of bound states cannot
exist and  indeed, it has been rigorously proved (see e.g.
\cite{AZGG}, \cite{BCDSSW}) that for sufficiently regular periodic
potentials (for singular, e.g. $\delta$-like potentials, the
situation might be different; see \cite{Be}, \cite{BDMN} and the
references therein) the spectrum is absolutely continuous in the
presence of a weak homogeneous time independent electric field so,
if Stark-Wannier ladders exist, they consist of resonances. The
issue remained controversial for decades and eventually settled
down in the affirmative at the rigorous level by using powerful
mathematical tools (for references and a detailed discussion see
sections IA, IV and VIA in \cite{N}). One of the key steps was the
proof in \cite{NN1} that one can define recurrently deformed bands
for which the interband transitions are smaller than any power of
the electric field strength. In its time independent form the
expansion method in \cite{NN1} has been considerably extended in
\cite{S}, \cite{GMS}. Considered initially as an interesting but
academic problem, the existence of Stark-Wannier ladders of
resonances was experimentally proved after the invention of
superlattices (see \cite{R} and the references therein) and even
more, found technological applications (see e.g. \cite{B}).

Since the time independent electric fields are (ideal) limits of
slowly varying in time electric fields it is naturally to try to
extend the whole analysis to slowly varying fields. At the
heuristic level one expects by an adiabatic argument that the
interband transitions are still small and one can hope to prove
the same type of result about the existence of almost invariant
deformed bands. Such a generalization was conjectured already in
\cite{NN1} and indeed, in \cite{NA1}, \cite{NA2} we developed a
similar theory as in the time independent case up to the second
order. Unfortunately, for higher orders the computations become
unmanageably complicated.

In this paper we shall develop a different procedure based on the
adiabatic expansion in \cite{NG} which allows us to push the
construction of the deformed bands for slowly varying in time
electric fields to arbitrary order.

The content of the paper is as follows: Section 2 contains a brief
review of the result in \cite{NN1} about time independent case,
the description of the problem and the main result. Section 3
contains the  construction of the  orthogonal projection on the
subspaces describing the deformed bands. Finally, Section 4
contains the proofs.

\section{The problem and the main result}

We begin with a short review of the main result in \cite{NN1}. For
simplicity we shall treat one-dimensional case, but the results
are valid for arbitrary dimensions.

The Hamiltonian describing one electron subjected to a periodic
potential and to a perturbation given by a homogeneous time
independent eletric field $E$ is:
\begin{equation}\label{H1}
H{^\varepsilon}=H_{0}+\varepsilon X_{0}
\end{equation}
where
\begin{equation}
\begin{array}{c}
  \varepsilon=-e E;\\
  H_{0}=-\frac{1}{2m}\frac{d^{2}}{dx^{2}}+V(x);\\
  V(x+na)=V(x)\\
\end{array}
\end{equation}
and {\it a} is the lattice constant.

The spectrum of $H_{0}$, $\sigma(H_{0})=\sigma_{0}$, is supposed
to have at least one isolated band $\sigma_{0}^{0}$ separated by
the rest of the spectrum:
$$
\sigma_{0}=\sigma_{0}^{0}\cup\sigma_{0}^{1}
$$
$$
dist(\sigma_{0}^{0},\sigma_{0}^{1})=d>0
$$

The mathematical difficulty of the problem comes from the fact
that even for low values of the electric field $E$, the potential
energy goes to infinity at large distances and the ordinary
perturbation theory cannot be applied. The Hamiltonian of the
perturbed system can be written in the following form:
$$
H^{\varepsilon}=P_{0}H^{\varepsilon}P_{0}+(1-P_{0})H^{\varepsilon}(1-P_{0})+
\left( P_{0}H^{\varepsilon}(1-P_{0})+h.c.\right)
$$
where $P_{0}$ is the orthogonal projection on the  subspaces of
states corresponding to the isolated band $\sigma_{0}^{0}$ of
$H_{0}$. As already remarked by Callaway \cite{C1}, \cite{C2}, the
one-band Hamiltonian $P_{0}H^{\varepsilon}P_{0}$ has a discrete
spectrum called Stark-Wannier ladder of the form
$\alpha+\varepsilon ak$, where $\alpha$ is a constant, $a$ the
crystal constant and $k$ an integer. As for in band dynamics, the
electron  is not continuous accelerated, but will undergo a
periodic motion in $k$-space caused by the Bragg reflections at
the boundary of the Brillouin zone, having the period
$T=\frac{2\pi}{\varepsilon a}$. This oscilatory motion in
$k$-space, accompanied by a periodic motion in the real space is
termed Bloch oscillations. The main issue was whether or not this
picture is washed out by the interband coupling $(
P_{0}H^{\varepsilon}(1-P_{0})+h.c.)$. Wannier \cite{W1}, \cite{W2}
argued that one can redefine the bands of $H_{0}$ so that the
one-band Hamiltonian
$$
P^{\varepsilon}H^{\varepsilon}P^{\varepsilon}
$$
where $P^{\varepsilon}$ is the orthogonal projection on the
subspace of states corresponding to a deformed band, has again
discrete spectrum and the non-diagonal part vanishes,
$P^{\varepsilon}H^{\varepsilon}(1-P^{\varepsilon})+h.c.=0$, i.e.
the deformed bands are "closed" under the dynamics given by
$H^{\epsilon}$. Unfortunately, as discussed in the Introduction,
the existence of closed bands is ruled out by the fact that the
spectrum of $H^{\varepsilon}$ is absolutely continuous.

The main result in \cite{NN1} is a recurrent rigorous construction
of deformed bands $\sigma_{0}^{n}$ so that the interband coupling
although nonzero are small, i.e. if $P_{n}^{\varepsilon}$ is the
orthogonal projection on the subspace of states corresponding to
the deformed band, then
$$
P_{n}^{\varepsilon}H^{\varepsilon}(1-P_{n}^{\varepsilon})
$$
is of the order $\varepsilon^{n+1}, n=1,2,...$. This implies that
\begin{equation}\label{tranz1}
 \gamma_{n}(\varepsilon,t)=
\|(1-P_{n}^{\varepsilon})e^{-iH^{\varepsilon}t}P_{n}^{\varepsilon}\|\leq
b_{n}\varepsilon^{n+1}t
\end{equation}
Taking into account that $1-\gamma_{n}(\varepsilon,t)^{2}$ is a
lower bound for the probability of finding at time $t$ the
electron in a state corresponding to $\sigma_{0}^{n}$ if at $t=0$
the electron is with probability one in a state corresponding to
$\sigma_{0}^{n}$, it follows that for states corresponding to
$\sigma_{0}^{n}$ and time scales of order $t \simeq \varepsilon
^{-n}$, the dynamics generated by the full Hamiltonian
$H^{\varepsilon}$ is well approximated by the dynamics generated
by the one-band Hamiltonian
$P_{n}^{\varepsilon}H^{\varepsilon}P_{n}^{\varepsilon}$.

Coming back to our time dependent electric field problem, the
Hamiltonian of the system is
\begin{equation}\label{H1t}
H^{\varepsilon, \omega}(t)=H_{0}+\varepsilon X_{0}F(\omega t)
\end{equation}
with $F(u)$ and all its derivatives $F^{(n)}(u)$ bounded. The case
$F(u)=1$ is the one discussed above.

Heuristically, it is expected by an adiabatic argument that for
small $\omega$ the transitions caused by the time dependence of
the electric field are still small and one hope the same type of
result. More precisely, if $U^{\varepsilon,\omega}(t)$ is the
solution of the Schr\"{o}dinger equation
\begin{equation}
i\frac{dU^{\varepsilon, \omega}(t)}{dt}=H^{\varepsilon,
\omega}(t)U^{\varepsilon, \omega}(t)
\end{equation}
we are looking for an operator $P_{n}^{\varepsilon, \omega}(t)$,
$n=0,1,2,...$, $P_{0}^{\varepsilon, \omega}(t)=P_{0}$, so that the
interband transitions be bounded by
\begin{equation}\label{tranz2}
\gamma_{n}(\varepsilon, \omega,t) =\parallel(1-P_{n}^{\varepsilon,
\omega}(t))U^{\varepsilon, \omega}(t)P_{n}^{\varepsilon,
\omega}(t)\parallel \leq
t\varepsilon\sum_{\alpha=0}^{n}C_{\alpha}\varepsilon^{n-\alpha}\omega^{\alpha}
\end{equation}

A recurrent construction of $P_{n}^{\varepsilon, \omega}(t)$ such
that \eqref{tranz2} holds true is the main result of this paper.

We end up this section with a few remarks.

i. As expected, in the limit $\omega \rightarrow 0$ \eqref{tranz2}
 reduces to \eqref{tranz1}.

ii. $P_{n}^{\varepsilon, \omega}(t)$ is constructed out of
$H^{\varepsilon, \omega}(t)$ and its derivatives up to order $n$.

iii. As in the time independent electric field case \cite{NN1} the
smallness of interband transitions implies the validity of one
-band approximation on long time scales (of order
$\min_{\alpha=0,1,...n}(\varepsilon^{n-\alpha}\omega^{\alpha})^{-1}$,
$n=1,2,...$). However, since both $H^{\varepsilon, \omega}$ and
$P_{n}^{\varepsilon, \omega}$ depend on time, the analysis of the
one-band dynamics is more complicated than in the time independent
electric field case \cite{NN2} and is deferred to a future
publication.

iv. As already said in the Introduction, in \cite{NA1}, \cite{NA2}
we developed for the above Hamiltonian \eqref{H1t} a similar
theory as in the time independent case up to the second order.
More exactly, we redefined the deformed bands of $H_{0}$ and for
these deformed bands, in the second order theory the interband
transitions are bounded by
\begin{equation}
\gamma_{1}(\varepsilon, \omega,
t)\leq(C_{1}\varepsilon^{2}+C_{2}\varepsilon \cdot \omega)\mid
t\mid
\end{equation}
The recurrent procedure was not developed further to an arbitrary
order $n$, the higher order construction implying very laborious
calculations.

\section{Construction of $P_{n}^{\varepsilon, \omega}(t)$}

In the following we shall use a procedure based on the adiabatic
expansion theorem developed in \cite{NG}.

Unfortunately, the Hamiltonian of the problem \eqref{H1t} is not
of an adiabatic type. Moreover, in this problem we are dealing
with two small parameters $\varepsilon$ and $\omega$.

If we rescale
$$
\begin{array}{ccc}
  s=\varepsilon t; & \omega=\varepsilon a; & a-parameter \\
\end{array}
$$
the Schr\"{o}dinger equation becomes:
\begin{equation}\label{HA}
i\varepsilon\frac{dU^{\varepsilon}(s, a)}{ds}=H^{\varepsilon}(s,a)
U^{\varepsilon}(s, a)
\end{equation}

Defining
\begin{equation}\label{U}
U_{0}(s, a)\equiv e^{-iX_{0}G(s,a)}
\end{equation}
where
\begin{equation}\label{G}
G(s,a)=\int_{0}^{s}F(au)du
\end{equation}
and
\begin{equation}
W^{\varepsilon}(s,a)\equiv U_{0}^{\ast}(s,a)U^{\varepsilon}(s,a)
\end{equation}
the Schr\"{o}dinger equation becomes of the adiabatic form
\cite{NG}, but with an aditional parameter $a$:
\begin{equation}\label{W1}
 i\varepsilon\frac{dW^{\varepsilon}(s,a)}{ds}=
 \widetilde{H}_{0}(s,a)W^{\varepsilon}(s,a)
\end{equation}
where
\begin{equation}\label{tH0}
\widetilde{H}_{0}(s,a)=U_{0}^{\ast}(s,a)H_{0}U_{0}(s,a)
\end{equation}
has the same spectrum as $H_{0}$.

Now, in terms of $W^{\varepsilon}(s,a)$ the interband transitions
\eqref{tranz2} become \cite{NA1}:
\begin{equation}\label{T2}
\gamma_{n}(\varepsilon, \omega, t)=\gamma_{n}(\varepsilon,
s,a)\equiv
\parallel(1-\widetilde{P}_{n}^{\varepsilon }(s,a)W^{\varepsilon}(s,a)
\widetilde{P}_{n}^{\varepsilon}(0,a)\parallel
\end{equation}
where
\begin{equation}\label{PP}
\widetilde{P}_{n}^{\varepsilon}(s,a)=U_{0}^{\ast}(s,a)P_{n}^{\varepsilon,\omega}(t)U_{0}(s,a)
\end{equation}
have to be constructed. Once
$\widetilde{P}_{n}^{\varepsilon}(s,a)$ constructed,
$P_{n}^{\varepsilon,\omega}(t)$ are given by \eqref{PP}. At fixed
$a$, the construction of $\widetilde{P}_{n}^{\varepsilon}(s,a)$
follows closely the method in \cite{NG} but emphasizing the $a$
dependence.

We define the sequence $\widetilde{E}_{j}(s,a)$ by the recurrence
formula (see Lemma 1 in \cite{NG}):
\begin{equation}\label{E0}
\widetilde{E}_{0}(s,a)=\widetilde{P}_{0}(s,a)=
\frac{i}{2\pi}\oint_{\Gamma}\frac{1}{\widetilde{H}_{0}(s,a)-z}dz
=\frac{i}{2\pi}\oint_{\Gamma}\widetilde{R}_{0}(s,a;z)dz
\end{equation}
$$
\widetilde{E}_{j}(s,a) =
-\frac{1}{2\pi}\oint_{\Gamma}\widetilde{R}_{0}(s,a;z)
[(1-\widetilde{P}_{0}(s,a))\widetilde{E}_{j-1}^{(1)}(s,a)\widetilde{P}_{0}(s,a)-
h.c.]\widetilde{R}_{0}(s,a;z)dz+
$$
\begin{equation}\label{Ej}
+\widetilde{S}_{j}(s,a)-
2\widetilde{P}_{0}(s,a)\widetilde{S}_{j}(s,a)\widetilde{P}_{0}(s,a)
\end{equation}
where
\begin{equation}\label{Sj}
\widetilde{S}_{j}(s,a)=\sum_{m=1}^{j-1}\widetilde{E}_{m}(s,a)
\widetilde{E}_{j-m}(s,a)
\end{equation}
$$
\widetilde{E}_{j}^{(n)}(s,a)=\frac{d^{n}\widetilde{E}_{j}(s,a)}{ds^{n}}
$$
and $\Gamma$ is a contour enclosing the isolated band
$\sigma_{0}^{0}$. $\widetilde{E}_{j}(s,a)$ satisfy:
\begin{equation}\label{E1}
\widetilde{E}_{j}(s,a)=\sum_{m=0}^{j}\widetilde{E}_{m}(s,a)\widetilde{E}_{j-m}(s,a)
\end{equation}
\begin{equation}\label{E2}
i\widetilde{E}_{j-1}^{(1)}(s,a)=[\widetilde{H}_{0}(s,a),\widetilde{E}_{j}(s,a)]
\end{equation}
As a consequence of \eqref{E1}, \eqref{E2},
$T_{n}^{\varepsilon}(s,a)$, $n=0,1,2,...$ defined by:
\begin{equation}\label{Tn1}
T_{n}^{\varepsilon}(s,a)=\sum_{j=0}^{n}\widetilde{E}_{j}(s,a)\varepsilon^{j}
\end{equation}
have the properties:
\begin{equation}\label{Tn2}
i\varepsilon T_{n}^{{\varepsilon}(1)}-
[\widetilde{H}_{0}(s,a),T_{n}^{\varepsilon}(s,a)]=
i\widetilde{E}_{n}^{(1)}\varepsilon^{n+1}
\end{equation}
$$
\parallel\left(T_{n}^{\varepsilon}(s,a)\right)^{2}-T_{n}^{\varepsilon}(s,a)\parallel
\sim \mathcal{O}(\varepsilon^{n+1})
$$

Finally, following \cite{NG}, \cite{NG1} we construct  projection
operators $\widetilde{P}_{n}^{\varepsilon}(s,a)$ corresponding to
almost invariant subspaces describing the deformed bands:
$$
\widetilde{P}_{n}^{\varepsilon}(s,a)= \frac{i}{2\pi}\int_{\mid
z-1\mid=\frac{1}{2}}
\left(T_{n}^{\varepsilon}(s,a)-z\right)^{-1}dz=
$$
\begin{equation}
=T_{n}^{\varepsilon}(s,a)+
\left(T_{n}^{\varepsilon}(s,a)-\frac{1}{2}\right)\left\{
\left[1+4\left((T_{n}^{\varepsilon}(s,a))^{2}-T_{n}^{\varepsilon}(s,a)\right)\right]
^{-\frac{1}{2}}-1\right\}
\end{equation}

The crucial property of $\widetilde{P}_{n}^{\varepsilon}(s,a)$ is:
$$
i\varepsilon\widetilde{P}_{n}^{\varepsilon (1)}(s,a)-
\left[\widetilde{H}_{0}(s,a),\widetilde{P}_{n}^{\varepsilon}(s,a)\right]=
$$
\begin{equation}\label{Pn}
=-\varepsilon^{n+1}\frac{1}{2\pi}\int_{\mid
z-1\mid=\frac{1}{2}}\left(T_{n}^{\varepsilon}(s,a)-z\right)^{-1}\widetilde{E}_{n}^{(1)}(s,a)
\left(T_{n}^{\varepsilon}(s,a)-z\right)^{-1}dz
\end{equation}

Using the fact that
$(1-\widetilde{P}_{n}^{\varepsilon}(s,a))\widetilde{P}_{n}^{\varepsilon}(s,a)
=0$  and that
$\|(1-\widetilde{P}_{n}^{\varepsilon}(s,a))\|=\|W^{\varepsilon}(s,a)\|=1$
the interband transitions \eqref{T2} can be rewritten as:
$$
\gamma_{n}(\varepsilon,s,a)=\parallel\left(1-\widetilde{P}_{n}^{\varepsilon}(s,a)\right)
W^{\varepsilon}(s,a)\widetilde{P}_{n}^{\varepsilon}(0,a)W^{\varepsilon
\ast}(s,a)W^{\varepsilon}(s,a)\parallel =
$$
\begin{equation}\label{T3}
\parallel\left(1-\widetilde{P}_{n}^{\varepsilon}(s,a)\right)
\left[-\widetilde{P}_{n}^{\varepsilon}(s,a)+
W^{\varepsilon}(s,a)\widetilde{P}_{n}^{\varepsilon}(0,a)W^{\varepsilon
\ast}(s,a)\right]W^{\varepsilon}(s,a)\parallel \leq
\end{equation}
$$
\leq \|\widetilde{P}_{n}^{\varepsilon}(s,a)-
W^{\varepsilon}(s,a)\widetilde{P}_{n}^{\varepsilon}(0,a)W^{\varepsilon
\ast}(s,a)\|
$$

It remains to estimate the last norm in \eqref{T3}. The main point
is that in order
 to obtain
estimations of the form \eqref{tranz2} one has to control the $a$
dependence.

\section{Proofs}
 We begin with a preparatory result.
\begin{lemma}\label{L1}

$$
\parallel \widetilde{P}_{n}^{\varepsilon}(s,a)-
W^{\varepsilon}(s,a)\widetilde{P}_{n}^{\varepsilon}(0,a)W^{\varepsilon
\ast}(s,a)\parallel
$$
\begin{equation}\label{kato}
\leq\frac{1}{\varepsilon}\int_{0}^{s}\parallel i\varepsilon
\frac{d\widetilde{P}_{n}^{\varepsilon}(u,a)}{du}-\left[\widetilde{H}_{0}(u,a),
\widetilde{P}_{n}^{\varepsilon}(u,a)\right]\parallel du
\end{equation}
\end{lemma}

{\em Proof.} The proof is standard \cite{Ka},\cite{NG} but we give
it for completeness. Rewrite the l.h.s. of \eqref{kato} as:
$$
\widetilde{P}_{n}^{\varepsilon}(s,a)-
W^{\varepsilon}(s,a)\widetilde{P}_{n}^{\varepsilon}(0,a)W^{\varepsilon
\star}(s,a)=
$$
$$
=W^{\varepsilon}(s,a)\left[ W^{\varepsilon
\star}(s,a)\widetilde{P}_{n}^{\varepsilon}(s,a)W^{\varepsilon}(s,a)-
\widetilde{P}_{n}^{\varepsilon}(0,a) \right]W^{\varepsilon
\star}(s,a)
$$

Using \eqref{W1}, the equation satisfied by the function
$$
f(s,a)=W^{\varepsilon
\star}(s,a)\widetilde{P}_{n}^{\varepsilon}(s,a)W^{\varepsilon}(s,a)-
\widetilde{P}_{n}^{\varepsilon}(0,a)
$$
 is
$$
i\varepsilon \frac{df(s,a)}{ds}=W^{\varepsilon \ast}(s,a)\left \{
i\varepsilon\frac{d\widetilde{P}_{n}^{\varepsilon}(s,a)}{ds}
-\left[\widetilde{H}_{0}(s,a),\widetilde{P}_{n}^{\varepsilon}(s,a)\right]
\right\}W^{\varepsilon}(s,a)
$$
 The solution of this equation is
 $$
f(s,a)-f(0,a)=\frac{1}{i\varepsilon}\int_{0}^{s}W^{\varepsilon
\ast}(u,a)
\left\{i\varepsilon\frac{d\widetilde{P}_{n}^{\varepsilon}(u,a)}{ds}
-\left[\widetilde{H}_{0}(u,a),\widetilde{P}_{n}^{\varepsilon}(u,a)\right]
\right\}W^{\varepsilon}(u,a)du
 $$
 Since  $W^{\varepsilon}(s,a)$ is  unitary    and $f(0,a)=0$, Lemma
\ref{L1} results immediately. As a result \eqref{T3} becomes:
\begin{equation}\label{T4}
\gamma_{n}(\varepsilon,s,a)\leq \frac{1}{\varepsilon}\int_{0}^{s}
\parallel
i\varepsilon\frac{d\widetilde{P}_{n}^{\varepsilon}(u,a)}{du}-
\left[\widetilde{H}_{0}(u,a),\widetilde{P}_{n}^{\varepsilon}(u,a)\right]\parallel
du
\end{equation}

Now from \eqref{T4},  the property \eqref{Pn} of the projection
operators $\widetilde{P}_{n}^{\varepsilon}(s,a)$ and the fact that
(\cite{NG}, \cite{NG1})  $ \sup_{\mid
z-1\mid=\frac{1}{2}}\|\left(T_{n}^{^{\varepsilon}}(s,a)-z\right)^{-1}\|$
is bounded uniformly in $s$
 it results:
\begin{equation}\label{rate}
\gamma_{n}(\varepsilon,s,a) \leq const.\varepsilon^{n} sup_{0\leq
u \leq s}\parallel \widetilde{E}_{n}^{(1)}(u,a)\parallel\cdot s
\end{equation}
and what is left is to obtain estimations of
$\|\widetilde{E}_{n}^{(1)}(u,a)\parallel$.

In what follows
$\widetilde{R}_{0}(s,a;z)=(\widetilde{H}_{0}(s,a)-z)^{-1}$,
$R_{0}(z)=(H_0-z)^{-1}$ and $\Gamma$ a contour enclosing
$\sigma_{0}^{0}$. We shall prove first:

\begin{lemma}\label{L2}
\begin{equation}
\sup_{s\in \mathbb{R}, z\in \Gamma
}\parallel\widetilde{R}_{0}^{(n)}(s,a;z)\parallel\leq
\sum_{l=0}^{n-1}C_{l}a^{l}
\end{equation}
\end{lemma}
{\em Proof.} For $n=1,2,3$, by a direct calculation using
\eqref{U}, \eqref{G} and \eqref{tH0} one obtains:

$$
\widetilde{R}_{0}^{(1)}(s,a;z) =iF(as)U_{0}^{\ast}(s,a)
\left[X_{0},R_{0}(z)\right]U_{0}(s,a)
$$

$$
\widetilde{R}_{0}^{(2)}(s,a;z)=iaF^{(1)}(as) U_{0}^{\ast}(s,a)
\left[X_{0},R_{0}(z)\right]U_{0}(s,a)+
$$
$$
+F^{2}(as)U_{0}^{\ast}(s,a)\left
[\left[X_{0},R_{0}(z)\right],X_{0}\right] U_{0}(s,a)
$$

$$
\widetilde{R}_{0}^{(3)}(s,a;z)=ia^{2}F^{(2)}(as) U_{0}^{\ast}(s,a)
\left[X_{0},R_{0}(z)\right]U_{0}(s,a)+
$$
$$
+aF^{(1)}(as)F(as)U_{0}^{\ast}(s,a)\left
[\left[X_{0},R_{0}(z)\right],X_{0}\right] U_{0}(s,a)-
$$
$$
-iF^{3}(as)U_{0}^{\ast} \left[\left[\left[
X_{0},R_{0}(z)\right],X_{0}\right],X_{0}\right]
$$

In general, one can see recurrently that
$\widetilde{R}_{0}^{(n)}(s,a)$ is a polynomial of degree $n-1$ in
$a$ whose coefficients are products of $F^{k}(as)$,
$U_{0}^{\ast}(s,a)$, $U_{0}(s,a)$ and multiple commutators
$[[...[R_{0}(z),X_{0}],...,X_{0}]]$. Since all these factors (for
the multiple commutators see e.g. \cite{A}, \cite{NN1}) are
uniformly bounded in $a$, $s$ and $z$ the proof of lemma is
finished.

Finally the next lemma gives the necessary estimate of
$\|\widetilde{E}_{n}^{(1)}(u,a)\parallel$:
\begin{lemma}\label{L3}
\begin{equation}\label{Eja}
\parallel \widetilde{E}_{j}(s,a)\parallel\leq
\sum_{l=0}^{j-1}e_{l}a^{l}
\end{equation}
\begin{equation}\label{E1ja}
\parallel \widetilde{E}_{j}^{(1)}(s,a)\parallel\leq
\sum_{l=0}^{j}f_{l}a^{l}
\end{equation}
\end{lemma}

{\em Proof.} We shall prove by induction that
$\widetilde{E}_{j}(s,a)$ is a finite sum of terms, each term is a
multiple integral on $\Gamma$, the integrand being
\begin{equation}\label{R}
\prod_{k=1}^{m}\widetilde{R}_{0}^{(\alpha_{k})}(s,a;z)
\end{equation}
where
$$\alpha_{k}\geq 0$$
and
$$\sum_{k}\alpha_{k}=1$$
 In addition, $\widetilde{E}_{j}^{(1)}(s,a)$ have the same form
 with
$$
\sum_{k}\alpha_{k}=j+1
$$

For $j=0$ this is trivial since (see \eqref{E0}):
$$
\widetilde{E}_{0}(s,a)=\widetilde{P}_{0}(s,a)= \frac{i}{2\pi}
\oint_{\Gamma}\widetilde{R}_{0}(s,a;z)dz
$$
and
$$
\widetilde{E}_{0}^{(1)}(s,a)=\frac{i}{2\pi}\oint_{\Gamma}\frac{d\widetilde{R}_{0}(s,a;z)}{ds}dz
$$

Suppose that $\widetilde{E}_{j}(s,a)$ satisfies the induction
hypothesis and we want  to prove the same is true for $j+1$.

From \eqref{Ej} $\widetilde{E}_{j+1}(s,a)$ contains two types of
terms:

- The first type is a multiple integral of terms containing
$\widetilde{E}_{j}^{(1)}(s,a)$ and resolvents of
$\widetilde{H}_{0}(s,a)$. According to the induction hypothesis
the terms are of the form \eqref{R}

where
$$
\sum_{k}\alpha_{k}=j+1
$$

- the second type of terms contains $\widetilde{S}_{j+1}(s,a)$:
$$
\widetilde{S}_{j+1}(s,a)=\sum_{m=1}^{j}\widetilde{E}_{m}(s,a)\widetilde{E}_{j+1-m}(s,a)
$$
and again from the induction hypothesis they are  of the above
form \eqref{R} with
$$
\sum_{k}\alpha_{k}=m+j+1-m=j+1
$$

It results that $\widetilde{E}_{j+1}$ is a finite sum of terms,
each term being a multiple integral on $\Gamma$, with the
integrant of the form
$$
\prod_{k=1}^{m}\widetilde{R}_{0}^{(\alpha_{k})}(s,a;z)
$$
with
$$
\alpha_{k}\geq 0;\;\;\;\sum_{k}\alpha_{k}=j+1
$$

By the Leibnitz rule, the derivative
$\widetilde{E}_{j+1}^{(1)}(s,a)$ is of the same form, but with
$$
\sum_{k}\alpha_{k}=j+2
$$
This and Lemma \ref{L2} give \eqref{Eja} and \ref{E1ja} which
finishes the proof.

Plugging \eqref{E1ja} into \eqref{rate} one obtains that

$$
\gamma_{n}(\varepsilon,s,a)\leq \varepsilon^{n}s\sum_{k}C_{k}a^{k}
$$
and going back to the variables $t$ and taking into account that
$a= \frac{\omega}{\varepsilon}$ it results
$$
\gamma_{n}(\varepsilon, \omega, t)\leq \varepsilon t
\sum_{k=0}^{n}C_{k}\varepsilon^{n-k}\omega^{k}
$$
which is the desired result.

\vspace{1cm}

\noindent{\bf Acknowledgements.}

I would like to thank G. Nenciu for suggesting me the adiabatic
expansion formalism and for helpful discussions. This research was
supported by CNCSIS under Grant 905-6/2007.

\end{document}